\documentstyle[preprint,aps,prd,psfig]{revtex}

\newcommand{\AmS}{{\protect\the\textfont2
  A\kern-.1667em\lower.5ex\hbox{M}\kern-.125emS}}
\newcommand{\be}{\begin{equation}}
\newcommand{\ee}{\end{equation}}
\newcommand{\ben}{\begin{eqnarray}}
\newcommand{\een}{\end{eqnarray}}
\newcommand{\nn}{\nonumber}
\newcommand{\tr}{{\rm Tr}}
\newcommand{\slas}[2]{{{#1}\hspace{-5pt}{/}}_{#2}}

\def\simgt{\rlap{\lower 3.5 pt\hbox{$\mathchar \sim$}}\raise 1pt \hbox {$>$}}
\def\simlt{\rlap{\lower 3.5 pt\hbox{$\mathchar \sim$}}\raise 1pt \hbox {$<$}}
\newcommand{\cont}{{\rm cont}}
\newcommand{\latt}{{\rm latt}}
\newcommand{\diag}{{\rm diag}}
\newcommand{\mix}{{\rm mix}}
\newcommand{\hk}{{\hat k}}
\newcommand{\csw}{{c_{\rm SW}}}
\newcommand{\lattint}{{\int_{-\pi}^\pi\frac{d^4 k}{\pi^2}}}
\newcommand{\cfactor}{{C_{B\hspace{-5pt}{/}\hspace{3pt}}}}
\newcommand{\rcovd}{{\stackrel{\rightarrow}{D\hspace{-8pt}{/}}}}
\newcommand{\lcovd}{{\stackrel{\leftarrow}{D\hspace{-8pt}{/}}}}

\begin{document}
\draft
\title{Perturbative Renormalization Factors of Baryon
Number Violating Operators for Improved Quark and Gauge Actions 
in Lattice QCD}

\author{Sinya~Aoki$^{\rm a}$,
        Yoshinobu~Kuramashi$^{\rm b}$
\thanks{On leave from Institute of Particle and Nuclear Studies,
        High Energy Accelerator Research Organization(KEK),
        Tsukuba, Ibaraki 305-0801, Japan},
        Tetsuya~Onogi$^{\rm c}$ and
        Naoto~Tsutsui$^{\rm c}$
       }
\address{$^a$Institute of Physics,
             University of Tsukuba,
             Tsukuba, Ibaraki 305-8571, Japan\\
         $^b$Department of Physics,
             Washington University,
             St. Louis, Missouri 63130, USA\\
         $^c$Department of Physics,
             Hiroshima University,
             Higashi-Hiroshima, Hiroshima 739-8526, Japan
        }

\date{\today}

\maketitle

\begin{abstract}
\vspace{-10mm}

We calculate one-loop renormalization factors of three-quark
operators, which appear in the low energy effective Lagrangian
of the nucleon decay, for $O(a)$-improved quark action
and gauge action including six-link loops.
This calculation is required to predict the hadronic
nucleon decay matrix elements in the continuum
regularization scheme using lattice QCD.
We present detailed numerical results of the one-loop coefficients
for general values of the clover coefficients 
employing the several improved gauge actions in the Symanzik approach and
in the Wilson's renormalization group approach.
The magnitudes of the one-loop coefficients for the
improved gauge actions show sizable reduction
compared to those for the plaquette action.

\end{abstract}

\section{Introduction}
 
While nucleon decay is one of the most exciting prediction
from grand unified theories (GUTs) with and without supersymmetry, 
none of the decay modes have been experimentally detected up to now.
Furthermore, the ongoing Super-Kamiokande experiment 
is now pushing the lower limit on the partial lifetimes 
of the nucleon by an order of magnitude from the previous
measurements. In principle this would give a strong constraint 
on (SUSY-)GUTs, however, the uncertainties in 
the theoretical prediction of lifetimes due to poor knowledge
of quantum effects at low energy such as hadron or SUSY scales 
obscure the direct impact of the experimental lifetime bound 
on the physics at the GUT scale. In particular, one of the main 
uncertainties has been found in the evaluation of the hadron matrix 
elements for the nucleon decays, for which various QCD models have 
given estimate differing by a factor of ten. Therefore precise 
determination of the nucleon decay matrix elements from first 
principles is required, for which Lattice QCD can play a crucial role.

Recently we carried out a model-independent calculation
of the nucleon decay matrix elements employing
the Wilson quark action and the plaquette gauge action
in the quenched approximation\cite{jlqcd_pd}. Although one naively expects 
the error in the discretization and quenching approximation
which are the two main systematic errors,
it would be desirable to reduce these unknown systematic errors 
for high precision calculation. As a step toward this goal we are required 
to reduce the scaling violation effects by improving quark and gauge 
actions.
In this article we present perturbative results for renormalization 
factors of baryon number violating operators for improved quark and gauge
actions: the $O(a)$-improved ``clover'' action originally
proposed by Sheikholeslami and Wohlert\cite{imp_q} and the gauge action
improved by addition of six link loops to the plaquette term
in the Symanzik approach\cite{imp_g_sym} 
and in the Wilson's renormalization group
approach\cite{imp_g_wil,imp_g_iwa}.
Values of the one-loop  coefficients of the renormalization
factors are numerically evaluated for combinations of general values 
of the clover coefficients in the quark action and some specific values
of the coefficients of the six-link loop terms in the gauge action.  

This paper is organized as follows. In Sec.\ref{sec:feynmanrule} 
we give the improved quark and gauge actions on the lattice 
and their Feynman rules relevant for their calculation.
Our calculational procedure of the renormalization factors
for the baryon number violating operators is described in
Sec.\ref{sec:calculation}, where we present expressions and numerical values
for the one-loop coefficients of the renormalization factors.
Our conclusions are summarized in Sec.\ref{sec:conclusion}.

The physical quantities are expressed in lattice units, and the
lattice spacing $a$ is suppressed unless necessary.
Throughout this paper we use the same notation for quantities defined
on the lattice and their counterparts in the continuum.
In case of any possibility of confusion, however, we shall make
a clear distinction between them.

\section{Actions and Feynman rules}
\label{sec:feynmanrule}

For the gauge  action we consider the following
general form including the standard plaquette term 
and six-link loop terms:
\be
S_{\rm gauge}=\frac{1}{g^2}\left\{ c_0\sum_{\rm plaquette}\tr U_{pl}
+c_1\sum_{\rm rectangle}\tr U_{rtg}
+c_2\sum_{\rm chair}\tr U_{chr}
+c_3\sum_{\rm parallelogram} \tr U_{plg} \right\}
\label{eq:action_g}
\ee
with the normalization condition
\be
c_0+8c_1+16c_2+8c_3=1,
\ee
where six-link loops are $1\times 2$ rectangle,
a bent $1\times 2$ rectangle (chair) and a three-dimensional
parallelogram obtained by multiplying the link variables
\be
U_{n,\mu}={\rm exp}\left(ig\sum_a T^a 
A_\mu^a(n+{\hat \mu}/{2})\right).
\ee
The free gluon propagator is obtained in Ref.\cite{imp_g_sym}:
\ben
D_{\mu\nu}(k)=\frac{1}{(\hk^2)^2}\left[
(1-A_{\mu\nu})\hk_\mu\hk_\nu+
\delta_{\mu\nu}\sum_\sigma\hk_\sigma^2 A_{\nu\sigma}\right]
\een
with 
\ben
\hk_\mu&=&2{\rm sin}\left(\frac{k_\mu}{2}\right),\\
\hk^2&=&\sum_{\mu=1}^{4}\hk_\mu^2.
\een
The matrix $A_{\mu\nu}$ satisfies
\ben
&({\rm i})& A_{\mu\mu}=0\;\;\; {\rm for}\;\;{\rm all}\;\; \mu, \\
&({\rm ii})& A_{\mu\nu}=A_{\nu\mu}, \\
&({\rm iii})& A_{\mu\nu}(k)=A_{\mu\nu}(-k). \\
&({\rm iv})& A_{\mu\nu}(0)=1\;\;\; {\rm for}\;\; \mu\ne\nu,
\een
and its expression is given by
\ben
A_{\mu\nu}(k)&=&\frac{1}{\Delta_4}
\left[(\hk^2-\hk_\nu^2)(q_{\mu\rho}q_{\mu\tau}\hk_\mu^2
+q_{\mu\rho}q_{\rho\tau}\hk_\rho^2
+q_{\mu\tau}q_{\rho\tau}\hk_\tau^2)\right. \nn\\
&&\left.+(\hk^2-\hk_\mu^2)(q_{\nu\rho}q_{\nu\tau}\hk_\nu^2
+q_{\nu\rho}q_{\rho\tau}\hk_\rho^2
+q_{\nu\tau}q_{\rho\tau}\hk_\tau^2)\right. \nn\\
&&\left.+q_{\mu\rho}q_{\nu\tau}(\hk_\mu^2+\hk_\rho^2)(\hk_\nu^2+\hk_\tau^2) 
+q_{\mu\tau}q_{\nu\rho}(\hk_\mu^2+\hk_\tau^2)(\hk_\nu^2+\hk_\rho^2)\right. \nn\\
&&\left.-q_{\mu\nu}q_{\rho\tau}(\hk_\rho^2+\hk_\tau^2)^2
-(q_{\mu\rho}q_{\nu\rho}+q_{\mu\tau}q_{\nu\tau})\hk_\rho^2\hk_\tau^2\right. \nn\\
&&\left.-q_{\mu\nu}(q_{\mu\rho}\hk_\mu^2\hk_\tau^2
           +q_{\mu\tau}\hk_\mu^2\hk_\rho^2
           +q_{\nu\rho}\hk_\nu^2\hk_\tau^2
           +q_{\nu\tau}\hk_\nu^2\hk_\rho^2)\right],
\label{eq:matrix_a}
\een
with $\mu\ne\nu\ne\rho\ne\tau$ the Lorentz indices. 
$q_{\mu\nu}$ and $\Delta_4$  are written as
\ben
q_{\mu\nu}&=&(1-\delta_{\mu\nu})
\left[1-(c_1-c_2-c_3)(\hk_\mu^2+\hk_\nu^2)-(c_2+c_3)\hk^2\right], \\
\Delta_4&=&\sum_\mu \hk_\mu^4\prod_{\nu\ne\mu}q_{\nu\mu}
+\sum_{\mu >\nu,\rho >\tau,\{\rho,\tau\}\cap\{\mu,\nu\}=\emptyset}
\hk_\mu^2\hk_\nu^2q_{\mu\nu}(q_{\mu\rho}q_{\nu\tau}
                            +q_{\mu\tau}q_{\nu\rho}).
\een
In the case of the standard plaquette action,
the matrix $A_{\mu\nu}$ is simplified as
\be
A_{\mu\nu}^{\rm plaquette}=1-\delta_{\mu\nu}.
\ee
 
For the quark action we consider the $O(a)$-improved quark action:
\ben
S_{\rm quark}&=&
\sum_n\frac{1}{2}\sum_\mu
\left\{{\bar \psi}_n(-r+\gamma_\mu)U_{n,\mu}\psi_{n+{\hat \mu}}
      +{\bar \psi}_n(-r-\gamma_\mu)U^\dagger_{n-{\hat \mu},\mu}
       \psi_{n-{\hat \mu}}\right\}
+(m_0+4r)\sum_n{\bar \psi}_n\psi_n \nn\\
&&-\csw\sum_n\sum_{\mu,\nu}ig\frac{r}{4}
{\bar \psi}_n\sigma_{\mu\nu}P_{\mu\nu}(n)\psi_n,
\label{eq:action_q}
\een
where we define the Euclidean gamma matrices in terms of
the Minkowski matrices in the Bjorken-Drell convention:
$\gamma_j=-i\gamma_{BD}^j$ $(j=1,2,3)$, 
$\gamma_4=\gamma_{BD}^0$,
$\gamma_5=\gamma_{BD}^5$ and 
$\sigma_{\mu\nu}=\frac{1}{2}[\gamma_\mu,\gamma_\nu]$.
The field strength $P_{\mu\nu}$ in the ``clover'' term
is given by
\ben 
P_{\mu\nu}(n)&=&\frac{1}{4}\sum_{i=1}^{4}\frac{1}{2ig}
\left(U_i(n)-U_i^\dagger(n)\right), \\
U_1(n)&=&U_{n,\mu}U_{n+{\hat \mu},\nu}
         U^\dagger_{n+{\hat \nu},\mu}U^\dagger_{n,\nu}, \\
U_2(n)&=&U_{n,\nu}U^\dagger_{n-{\hat \mu}+{\hat \nu},\mu}
         U^\dagger_{n-{\hat \mu},\nu}U_{n-{\hat \mu},\mu}, \\
U_3(n)&=&U^\dagger_{n-{\hat \mu},\mu}U^\dagger_{n-{\hat \mu}-{\hat \nu},\nu}
         U_{n-{\hat \mu}-{\hat \nu},\mu}U_{n-{\hat \nu},\nu}, \\
U_4(n)&=&U^\dagger_{n-{\hat \nu},\nu}U_{n-{\hat \nu},\mu}
         U_{n+{\hat \mu}-{\hat \nu},\nu}U^\dagger_{n,\mu}.
\een

From the quark action (\ref{eq:action_q}) we obtain
the free quark propagator 
\be
S_q^{-1}(p)=i\sum_\mu \gamma_\mu {\rm sin}(k_\mu)+m_0+
r\sum_\mu(1-{\rm cos}(p_\mu)).
\ee
In order to calculate renormalization factors of the baryon number
violating operators up to one-loop level, we need 
the following vertices,
\ben
V^a_{1\mu}(p,q)=-igT^a\left\{
\gamma_\mu{\rm cos}\left(\frac{p_\mu+q_\mu}{2}\right)
-ir{\rm sin}\left(\frac{p_\mu+q_\mu}{2}\right)\right\}, 
\label{eq:vertex_qg}\\
V^a_{c1\mu}(p,q)=-gT^a\csw\frac{r}{2}\left(
\sum_\nu \sigma_{\mu\nu}{\rm sin}(p_\nu-q_\nu)\right)
{\rm cos}\left(\frac{p_\mu-q_\mu}{2}\right)
\label{eq:vertex_c}
\een
with $p_\mu$ incoming quark momentum and $q_\mu$ outgoing
quark momentum.
The first vertex originates from 
the Wilson quark action and the second one is the interaction
due to the clover term. In the present calculation
of the vertex corrections for the baryon number violating operators,
the two-gluon vertices with quarks give no contribution.

We should note that the baryon number violating operators
contain a charge conjugated field, whose
action is obtained from eq.(\ref{eq:action_q}) with the replacement of
\be
T^a\longrightarrow -(T^a)^T,
\ee
where the superscript $T$ means the transposed matrix.
According to this change, the Feynman rule of the quark-gluon
vertices in eqs.(\ref{eq:vertex_qg}) and (\ref{eq:vertex_c}) 
should be modified
for the charge conjugated field.

\section{Renormalization factors for baryon number violating operators}
\label{sec:calculation}
\subsection{Calculational procedure}

We consider the following baryon number violating operators
in the continuum and on the lattice:
\ben
\left({\cal O}_{X,Y}^\cont\right)_\delta&=&\epsilon^{abc}
\left[({\bar \psi}_1^c)^a \Gamma_X (\psi_2)^b\right]
\left[ \Gamma_Y (\psi_3)^c \right]_\delta, 
\label{eq:op_cont}\\
\left({\cal O}_{X,Y}^\latt\right)_\delta&=&\epsilon^{abc}
\left[\left\{1+rm_0(1-z)\right\}({\bar \psi}_1^c)^a \Gamma_X (\psi_2)^b
\right.\nn\\
&&\left. +z\frac{r}{2}\left\{
({\bar \psi}_1^c\lcovd)^a\Gamma_X(\psi_2)^b-
({\bar \psi}_1^c)^a\Gamma_X(\rcovd\psi_2)^b\right\} 
\right.\nn\\
&&\left.-z^2\frac{r^2}{4}({\bar \psi}_1^c\lcovd)^a\Gamma_X
(\rcovd\psi_2)^b \right]\nn\\
&&\times\left[\left\{1+\frac{r}{2}m_0(1-z)\right\}\Gamma_Y (\psi_3)^c  
-z\frac{r}{2}\Gamma_Y(\rcovd\psi_3)^c\right]_\delta
\label{eq:op_latt}
\een
with
\ben
\rcovd\psi_n&=&\frac{1}{2}\sum_\mu\gamma_\mu
\left(U_{n,\mu}\psi_{n+{\hat \mu}}
-U^\dagger_{n-{\hat \mu},\mu}\psi_{n-\hat \mu}\right),\\
{\bar \psi}_n^c\lcovd&=&\frac{1}{2}\sum_\mu
\left({\bar \psi}^c_{n+{\hat \mu}}U^T_{n,\mu}
-{\bar \psi}^c_{n-\hat \mu}U^\ast_{n-{\hat \mu},\mu}\right)\gamma_\mu,
\een
where ${\bar \psi}^c=\psi^T C$ with $C=\gamma_4\gamma_2$ is a 
charge conjugated field of $\psi$.  Dirac structures are
represented by $\Gamma_X\otimes\Gamma_Y=
P_R\otimes P_R, P_R\otimes P_L, P_L\otimes P_R, P_L\otimes P_L$
with the right- and left-handed projection 
operators $P_{R,L}=(1\pm\gamma_5)/2$.
The summation over repeated color indices $a,b,c$ is assumed. 
 
Ultraviolet divergences of composite operators are regularized by the
cutoff $a^{-1}$ in the lattice regularization scheme, while
this is achieved by a reduction of the space-time dimension from four
in some continuum regularization schemes, where we consider
the naive dimensional regularization (NDR) scheme and the
dimensional reduction (DRED) scheme.
Operators defined in different regularization schemes can be related  by
renormalization factors:
\be
{\cal O}_{X,Y}^{\cont}(\mu)=Z_{\diag}(\mu a){\cal O}_{X,Y}^{\latt}(a)
+Z_{\mix}{\tilde {\cal O}}_{X,Y}^{\latt}(a)
\label{eq:relation}
\ee
with $\mu$ the continuum renormalization scale.
The explicit chiral symmetry breaking
due to the Wilson term in the quark action (\ref{eq:action_q})
causes the mixing between operators with different chiral structures,
which is denoted by
${\tilde {\cal O}}_{X,Y}^{\latt}$.
  
Since QCD is a asymptotically free theory, $Z_{\diag,\mix}$ are
expected to be perturbatively calculable in terms of the
coupling constant $g^2/(16\pi^2)$ at high energy scales,
which gives the following expressions,
\ben
Z_\diag(\mu a)&=&1+\frac{g^2}{16\pi^2}\left[ 
\frac{3}{2}\left(-(1-z)r\Sigma_0-\frac{8}{3}\ln(\mu a)+\Delta_\psi\right)
+8\ln(\mu a)+\Delta_{V,\diag}\right],
\label{eq:zfactor_diag}\\
Z_\mix&=&\frac{g^2}{16\pi^2}\Delta_{V,\mix}, 
\label{eq:zfactor_mix}
\een
where $\Sigma_0$ denotes the additive mass renormalization 
on the lattice, $\Delta_\psi$ is a contribution from
the wavefunction and $\Delta_V$ is from the vertex function. 
$\Sigma_0$ and $\Delta_\psi$ are
obtained by calculating the continuum and lattice quark self-energies
$\Sigma^{\cont,\latt}$ and $\Delta_V$ is from the continuum and lattice
vertex functions $\Lambda^{\cont,\latt}$.
  
The quark self-energies in the continuum and on the lattice 
are defined through the inverse
full quark propagators:
\ben
(S_q^\cont)^{-1}(p)&=&i\slas{p}{}+m-\frac{g^2}{16\pi^2}\Sigma^\cont(p), \\
(S_q^\latt)^{-1}(p)&=&i\sum_\mu\gamma_\mu {\rm sin}(p_\mu)+m_0
+r\sum_\mu(1-{\rm cos}(p_\mu))
-\frac{g^2}{16\pi^2}\Sigma^\latt(p),
\een
where we consider massless quark.
The difference between $\Sigma^{\cont}$ and $\Sigma^{\latt}$,
which originates from the regularization scheme dependence
of the self-energy, gives
the quark wavefunction renormalization factor,
\be
-\frac{8}{3}\ln(\mu a)+\Delta_\psi=
\left.\frac{\partial \Sigma^\cont(p)}{i\partial \slas{p}{}}\right|_{p=0}
-\left.\frac{\partial \Sigma^\latt(p)}{i\partial \slas{p}{}}\right|_{p=0}.
\ee
The additive quark mass renormalization is expressed as
\be
m_0-\frac{g^2}{16\pi^2}\Sigma_0
\ee
with
\be
\Sigma_0=\Sigma^\latt(p=0).
\ee
Notice that the renormalization factor (\ref{eq:zfactor_mix})
is given for the case of $m_0={g^2}/({16\pi^2})\Sigma_0$, where 
we consider the renormalization for massless quark. 

The vertex functions up to one-loop level 
in the continuum and on the lattice 
are expressed in the following way,
\ben
\Lambda_{X,Y}^\cont&=&\epsilon^{abc}\Gamma_X\otimes\Gamma_Y+
\frac{g^2}{16\pi^2}\epsilon^{abc}\Gamma_X\otimes\Gamma_Y V_{X,Y}^\cont,\\
\Lambda_{X,Y}^\latt&=&\epsilon^{abc}\Gamma_X\otimes\Gamma_Y+
\frac{g^2}{16\pi^2}\epsilon^{abc}\left[\Gamma_X\otimes\Gamma_Y 
\left(V_{X,Y}^\latt
+z{V_{X,Y}^{\prime\latt}}
-z^2{V_{X,Y}^{\prime\prime\latt}}\right)\right. \nn\\
&&\left.+{\tilde \Gamma}_X\otimes{\tilde \Gamma}_Y 
\left({\tilde V}_{X,Y}^\latt 
+z{\tilde V}_{X,Y}^{\prime\latt}
-z^2{\tilde V}_{X,Y}^{\prime\prime\latt}\right)
\right],
\een
where the number of prime in the superscript of the lattice 
vertex corrections denotes the number of covariant derivative applied
to the quark fields at the vertex. 
${\tilde \Gamma}_X\otimes{\tilde \Gamma}_Y$ term represents 
the mixing contribution.
The difference between $\Lambda_{X,Y}^{\cont}$ and 
$\Lambda_{X,Y}^{\latt}$ leads to 
\ben
8\ln(\mu a)+\Delta_{V,\diag}&=&V_{X,Y}^\cont-\left(V_{X,Y}^\latt
+z V_{X,Y}^{\prime\latt}-z^2 V_{X,Y}^{\prime\prime\latt}\right),\\
\Delta_{V,\mix}&=&-\left({\tilde V}_{X,Y}^\latt
+z {\tilde V}_{X,Y}^{\prime\latt}
-z^2 {\tilde V}_{X,Y}^{\prime\prime\latt}\right).
\een 
  
We note that the lattice quark-self energy and the lattice vertex
corrections are general function of the clover coefficient $\csw$
in the quark action and the six-link loop parameters $c_{1,2,3}$
in the gauge action.
Calculation of $\Delta_\psi$ was already carried out in Ref.\cite{pt_2_cr}
employing the general values for $\csw$.
For $c_{1,2,3}$ they choose some specific values:
$c_1=-1/12,c_2=c_3=0$ in the tree-level Symanzik improvement,
$c_1=-0.252,c_2=0,c_3=-0.17$ suggested by Wilson based on
renormalization group improvement
and $c_1=-0.331,c_2=c_3=0$ and $c_1=-0.27,c_2+c_3=-0.04$ by
Iwasaki.
According to Ref.\cite{pt_2_cr} we evaluate $\Delta_V$ 
for general values of $\csw$ and for the specific values of $c_{1,2,3}$
that they employed.

\subsection{Vertex corrections}

We calculate the vertex corrections of the operators
in eqs.(\ref{eq:op_cont}) and (\ref{eq:op_latt}) 
in the Feynman gauge
employing the massless quarks and
and the massless charge conjugated quark with momenta
$p_1=p_2=p_3=0$ as external states. 
The infrared singularities are regularized by a fictitious
gluon mass $\lambda$ introduced in the gluon propagator.

\begin{figure}[h]
\centering{
\hskip -0.0cm
\psfig{file=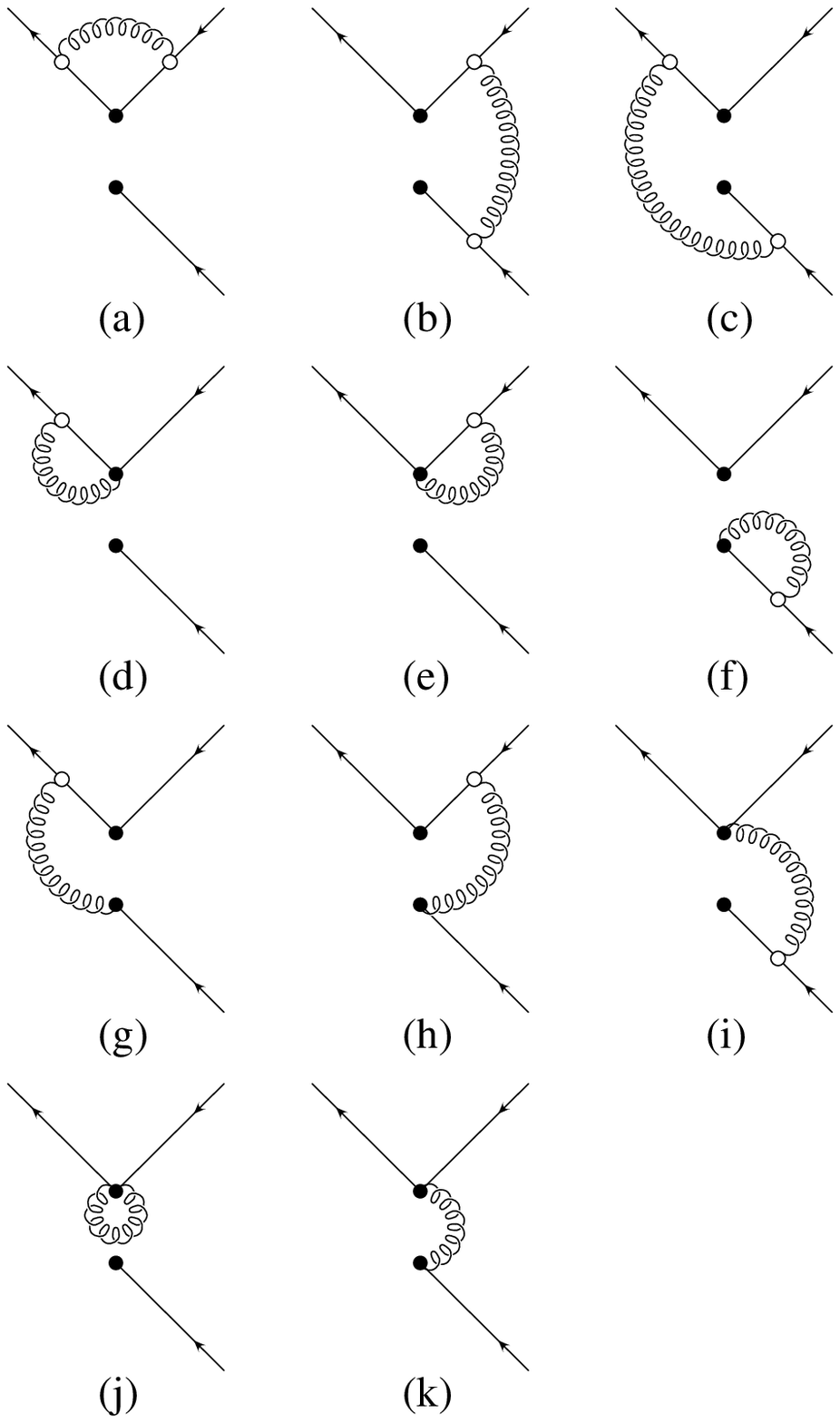,width=80mm,angle=0}
}
\caption{One-loop vertex corrections for the baryon number 
violating operator on the lattice.
Solid circle denotes the vertex. Open circle is for the
quark-gluon interaction either in eq.(\protect{\ref{eq:vertex_qg}}) 
or in eq.(\protect{\ref{eq:vertex_c}}).} 
\label{fig:ptdgm}
\end{figure}

One-loop vertex corrections on the lattice are illustrated
in Fig.\ref{fig:ptdgm}.
We find that the lattice vertex corrections 
is a second polynomial function of the clover coefficients $\csw$.
The relevant diagrams for $V_{X,Y}^\latt$ and ${\tilde V}_{X,Y}^\latt$ are 
Figs.\ref{fig:ptdgm}(a)$-$(i),
the sum of which gives
\ben
&&\Gamma_{R/L}\otimes\Gamma_Y V_{R/L,Y}
+{\tilde \Gamma}_{R/L}\otimes{\tilde \Gamma}_Y 
{\tilde V}_{R/L,Y} \nn\\
&=&\Gamma_{R/L}\otimes\Gamma_Y
\left(\cfactor 6\ln\left|\frac{1}{\lambda^2 a^2}\right|+
\sum_{i=0,1,2}\csw^{(i)}v_\diag^{(i)}\right) \nn\\
&&+\left(\Gamma_{L/R}\otimes\Gamma_Y\pm
\frac{1}{4}\sum_{\mu}\gamma_\mu\gamma_5\otimes\Gamma_Y\gamma_\mu \right)
\sum_{i=0,1,2}\csw^{(i)}v_\mix^{(i)}
\een
with $\cfactor=(N+1)/(2N)$ in the SU($N$) group. 
The explicit forms of $v_\diag^{(i)}$ and $v_\mix^{(i)}$ are given  by
\ben
v_\diag^{(0)}&=&\cfactor\lattint\left[\frac{1}{F_0^2G_0}\left\{
48(\Delta_3+r^2\Delta_1^2)^2+8I_a+16I_b\right\}
-\theta(\pi^2-k^2)\frac{6}{(k^2)^2}\right] \nn\\
&&+\cfactor 6\ln|\pi^2|, \\
v_\diag^{(1)}&=&\cfactor\lattint\left[\frac{1}{F_0^2G_0}\left\{
8r^2\Delta_1 I_a\right\}\right], \\
v_\diag^{(2)}&=&\cfactor\lattint\left[\frac{1}{F_0^2G_0}\left\{
4r^4\Delta_1^2 I_a\right\}\right], \\
v_\mix^{(0)}&=&\cfactor\lattint\left[\frac{1}{F_0^2G_0}\left\{
16r^2\Delta_1^2(\Delta_3-4\Delta^\mu_{1,0})\right\}\right], \\
v_\mix^{(1)}&=&\cfactor\lattint\left[\frac{1}{F_0^2G_0}\left\{
8r^2\Delta_1 I_a\right\}\right], \\
v_\mix^{(2)}&=&\cfactor\lattint\left[\frac{1}{F_0^2G_0}\left\{
-4r^2\Delta_3 I_a\right\}\right], 
\een
where 
\ben
F_0&=&\sum_\mu {\rm sin}^2(k_\mu)+\frac{r^2}{4}(\hk^2)^2, \\
G_0&=&(\hk^2)^2, \\
I_a&=&\Delta_{1,1}^\mu-4\Delta_3^2
+(16\Delta_3-4{\rm sin}^2(k_\mu))\Delta_{1,0}^\mu,\\
I_b&=&-\Delta_{1,1}^\mu+\Delta_3^2
+4(-\Delta_3+{\rm sin}^2(k_\mu))\Delta_{1,0}^\mu,\\
\Delta_1&=&\frac{1}{4}\hk^2, \\
\Delta_3&=&\frac{1}{4}\sum_\mu {\rm sin}^2(k_\mu), \\
\Delta_{1,1}^\mu&=&\sum_\nu(\delta_{\mu\nu}+A_{\mu\nu})
{\rm sin}^2(k_\mu) {\rm sin}^2(k_\nu), \\
\Delta_{1,0}^\mu&=&\sum_\nu(\delta_{\mu\nu}+A_{\mu\nu})
{\rm cos}^2\left(\frac{k_\mu}{2}\right) 
{\rm sin}^2\left(\frac{k_\nu}{2}\right).
\een
We do not take the sum over the index $\mu$ for 
$\Delta_{1,1}^\mu$ and $\Delta_{1,0}^\mu$.
In a similar way  we obtain the expressions of
$V^{\prime\latt}_{X,Y}$ and ${\tilde V}^{\prime\latt}_{X,Y}$
from Figs.\ref{fig:ptdgm}(a)$-$(i):
\ben
&&\Gamma_{R/L}\otimes\Gamma_Y V_{R/L,Y}^{\prime}
+{\tilde \Gamma}_{R/L}\otimes{\tilde \Gamma}_Y 
{\tilde V}_{R/L,Y}^{\prime} \nn\\
&=&\Gamma_{R/L}\otimes\Gamma_Y\sum_{i=0,1,2}\csw^{(i)}v_\diag^{\prime(i)} \nn\\
&&+\left(\Gamma_{L/R}\otimes\Gamma_Y\pm
\frac{1}{4}\sum_{\mu}\gamma_\mu\gamma_5\otimes\Gamma_Y\gamma_\mu \right)
\sum_{i=0,1,2}\csw^{(i)}v_\mix^{\prime(i)},
\een
where
\ben
v_\diag^{\prime(0)}&=&\cfactor\lattint\left[\frac{1}{F_0^2G_0}\left\{
16r^2\Delta_1 I_a+32r^2\Delta_1 I_b\right\}\right] \nn\\
&&+C_F\lattint\left[\frac{1}{F_0G_0}\left\{
12r^2\Delta_1(\Delta_3-4\Delta^\mu_{1,0})\right\}\right], \\
v_\diag^{\prime(1)}&=&\cfactor\lattint\left[\frac{1}{F_0^2G_0}\left\{
-8(r^2\Delta_3-r^4\Delta_1^2) I_a\right\}\right] \nn\\ 
&&+C_F\lattint\left[\frac{1}{F_0G_0}\left\{
3r^2 I_a\right\}\right], \\
v_\diag^{\prime(2)}&=&\cfactor\lattint\left[\frac{1}{F_0^2G_0}\left\{
-8r^4\Delta_1\Delta_3 I_a\right\}\right], \\
v_\mix^{\prime(0)}&=&\cfactor\lattint\left[\frac{1}{F_0^2G_0}\left\{
-32r^2\Delta_1\Delta_3(\Delta_3-4\Delta^\mu_{1,0})\right\} \right.\nn\\
&&\left.+\frac{1}{F_0G_0}\left\{
8r^2\Delta_1(\Delta_3-4\Delta^\mu_{1,0})\right\}\right], \\
v_\mix^{\prime(1)}&=&\cfactor\lattint\left[\frac{1}{F_0^2G_0}\left\{
-8(r^2\Delta_3-r^4\Delta_1^2) I_a\right\}
+\frac{1}{F_0G_0}\left\{2r^2 I_a\right\}\right], \\
v_\mix^{\prime(2)}&=&\cfactor\lattint\left[\frac{1}{F_0^2G_0}\left\{
-8r^4\Delta_1\Delta_3 I_a\right\}\right]
\een
with $C_F=(N^2-1)/(2N)$ in the SU($N$) group. 
The sum of Figs.\ref{fig:ptdgm}(a)$-$(k), 
which include the tadpole diagrams
at the vertex, yields $V^{\prime\prime\latt}_{X,Y}$ and 
${\tilde V}^{\prime\prime\latt}_{X,Y}$, 
\ben
&&\Gamma_{R/L}\otimes\Gamma_Y V_{R/L,Y}^{\prime\prime}
+{\tilde \Gamma}_{R/L}\otimes{\tilde \Gamma}_Y 
{\tilde V}_{R/L,Y}^{\prime\prime} \nn\\
&=&\Gamma_{R/L}\otimes\Gamma_Y\sum_{i=0,1,2}
\csw^{(i)}v_\diag^{\prime\prime(i)} \nn\\
&&+\left(\Gamma_{L/R}\otimes\Gamma_Y\pm
\frac{1}{4}\sum_{\mu}\gamma_\mu\gamma_5\otimes\Gamma_Y\gamma_\mu \right)
\sum_{i=0,1,2}\csw^{(i)}v_\mix^{\prime\prime(i)},
\een
where
\ben
v_\diag^{\prime\prime(0)}&=&\cfactor\lattint\left[\frac{1}{F_0^2G_0}\left\{
-8r^4\Delta_1^2 I_a-16r^4\Delta_1^2 I_b\right\}\right], \\
v_\diag^{\prime\prime(1)}&=&\cfactor\lattint\left[\frac{1}{F_0^2G_0}\left\{
8r^4\Delta_1\Delta_3 I_a\right\}\right], \\
v_\diag^{\prime\prime(2)}&=&\cfactor\lattint\left[\frac{1}{F_0^2G_0}\left\{
-4r^4\Delta_3^2 I_a\right\}\right], \\
v_\mix^{\prime\prime(0)}&=&\cfactor\lattint\left[\frac{1}{F_0^2G_0}\left\{
4r^2(16\Delta_3^2\Delta_{1,0}^\mu+8r^2\Delta_1^2\Delta_3^2
+4r^4\Delta_1^4\Delta_3)\right\} \right.\nn\\
&&\left.+\frac{1}{F_0G_0}\left\{
-8r^2(4\Delta_3\Delta_{1,0}^\mu+r^2\Delta_1^2\Delta_3)\right\}
+\frac{1}{G_0}\left\{
4r^2\Delta^\mu_{1,0}\right\}\right], \\
v_\mix^{\prime\prime(1)}&=&\cfactor\lattint\left[\frac{1}{F_0^2G_0}\left\{
8r^4\Delta_1\Delta_3 I_a\right\}
+\frac{1}{F_0G_0}\left\{-2r^4 \Delta_1 I_a\right\}\right], \\
v_\mix^{\prime\prime(2)}&=&\cfactor\lattint\left[\frac{1}{F_0^2G_0}\left\{
4r^6\Delta_1^2\Delta_3 I_a\right\}\right].
\een

We present numerical values of
$v_{\diag,\mix}^{(i)}$ in Table~\ref{tab:v}, 
$v_{\diag,\mix}^{\prime(i)}$ in Table~\ref{tab:vp}
and $v_{\diag,\mix}^{\prime\prime(i)}$ in Table~\ref{tab:vpp}, which
are evaluated with $r=1$ using the Monte Carlo integration routine
BASES\cite{bases} for specific values of $c_{1}$ and $c_2+c_3$.
The numerical accuracy is better than $0.01\%$.
Numerical values for $v_{\diag,\mix}^{(i)}$, 
$v_{\diag,\mix}^{\prime(i)}$ and $v_{\diag,\mix}^{\prime\prime(i)}$
can be also obtained by using 
the results for vertex corrections of bilinear operators in Ref.\cite{pt_2_cr},
in which the numerical values are evaluated in a different way. 
This is used as a check of our calculation. 
We note that a special case of $\csw=0$ and $c_1=c_2+c_3=0$
represents combination of the Wilson quark action and the 
plaquette gauge action, for which perturbative renormalization factors
for the baryon number violating operators has been 
already calculated\cite{pt_3_wp,jlqcd_pd}.
Comparison of our results with theirs gives us another check of
our calculation.
Numerical values in Tables~\ref{tab:v}, \ref{tab:vp} and \ref{tab:vpp}
show that the one-loop coefficients 
in the vertex corrections diminishes by $10-20\%$ 
for the tree-level Symanzik action
compared to those for the plaquette action.
Further reduction of the magnitude is observed 
for the renormalization group improved actions.
These features are also found 
in the case of bilinear operators\cite{pt_2_cr}.

In the continuum, the vertex correction at one-loop level is
expressed as
\be
V_{R/L,Y}=
\cfactor 6\ln\left|\frac{\mu^2}{\lambda^2}\right|+v_\diag.
\ee
The finite constant $v_\diag$ is given by
\ben
v_\diag^{\rm NDR} = \frac{8}{3}, \\
v_\diag^{\rm DRED} = 4
\een
for the NDR and DRED schemes with $\overline {\rm MS}$ subtraction.

The vertex corrections on the lattice give the expression of 
${\tilde {\cal O}}_{X,Y}^{\latt}$ in eq.(\ref{eq:relation}),
\be
{\tilde {\cal O}}_{R/L,Y}^{\latt}(a)=\epsilon^{abc}
\left[({\bar \psi}_1^c)^a \Gamma_{L/R} (\psi_2)^b\right]
\left[\Gamma_Y (\psi_3)^c\right]  
\pm\frac{1}{4}\sum_{\mu}
\epsilon^{abc}\left[({\bar \psi}_1^c)^a \gamma_\mu\gamma_5 (\psi_2)^b\right]
\left[\Gamma_Y\gamma_\mu(\psi_3)^c\right].  
\ee
Comparing the results for the vertex corrections in the continuum and 
on the lattice, we obtain the vertex correction 
components in the renormalization factors 
of eqs.(\ref{eq:zfactor_diag}) and (\ref{eq:zfactor_mix}),
\ben
\Delta_{V,\diag}^{\rm NDR,DRED}&=&v_\diag^{\rm NDR,DRED}-
\sum_{i=0,1,2}\csw^{(i)}(v_\diag^{(i)}+z v_\diag^{\prime(i)}
-z^2 v_\diag^{\prime\prime(i)}), \\
\Delta_{V,\mix}&=&
-\sum_{i=0,1,2}\csw^{(i)}(v_\mix^{(i)}+z v_\mix^{\prime(i)}
-z^2 v_\mix^{\prime\prime(i)}),
\een
where $\Delta_{V,\mix}$ is independent of the renormalization scheme
in the continuum.
To obtain the diagonal part of the renormalization factor 
in eq.(\ref{eq:zfactor_diag}), we also need the wavefunction
component $\Delta_\psi$. This quantity is already evaluated in
Ref.\cite{pt_2_cr} employing the NDR scheme in the continuum, 
where $C_F z_\psi$ in their notation corresponds to
our $\Delta_\psi^{\rm NDR}$. We note that 
$\Delta_\psi^{\rm DRED}$ can be
obtained from $\Delta_\psi^{\rm NDR}$ by
\be
\Delta_\psi^{\rm DRED}=\Delta_\psi^{\rm NDR}-\frac{4}{3}
\ee
For the mixing part of the
renormalization factor in eq.(\ref{eq:zfactor_mix}),  
$\Delta_\psi$ has no contribution.

\section{Conclusions}
\label{sec:conclusion}

In this paper we have calculated the one-loop contributions 
for the renormalization factors of the three-quark operators 
employing the improved quark and gauge actions.
Detailed numerical values of the one-loop coefficients
are presented for general values of the clover coefficients and 
for some specific values of $c_1$, $c_2$ and $c_3$ in the
gauge action improved by the Symanzik and 
renormalization group approaches.
We find that the magnitude of the one-loop coefficients 
are considerably reduced  
for the improved gauge actions
compared to those for the plaquette action, which is
a desirable feature in the practical implementation of
lattice QCD.

\acknowledgements

This work is supported in part by the Grants-in-Aid of
the Ministry of Education (No. 10740125).  
One of us (Y.K.) is supported by the JSPS Fellowship.



\newpage

\begin{table}[h]
\begin{center}
\caption{\label{tab:v} Finite parts $v_\diag$ and $v_\mix$ for
vertex correction of the baryon number violating operators.
Coefficients of the term $\csw^n$ $(n=0,1,2)$ are given in the 
column marked as $(n)$. } 
\begin{tabular}{ll|lll|lll}
\multicolumn{2}{c}{gauge action} & 
\multicolumn{3}{c}{$v_\diag$} & 
\multicolumn{3}{c}{$v_\mix$} \\
$c_1$ & $c_2+c_3$ & $(0)$ & $(1)$ & $(2)$ & $(0)$ & $(1)$ & $(2)$ \\
\hline
$0$      & $0$      & $11.084$  & $3.3282$ & $1.1498$   
                    & $-3.2137$ & $3.3282$ & $-1.1386$ \\  
$-1/12$  & $0$      & $10.743$  & $2.9496$ & $0.95371$   
                    & $-2.7085$ & $2.9496$ & $-1.0372$ \\  
$-0.331$ & $0$      & $10.052$  & $2.3000$ & $0.65335$   
                    & $-1.9381$ & $2.3000$ & $-0.84886$ \\  
$-0.27$  & $-0.04$  & $10.140$  & $2.3654$ & $0.67777$   
                    & $-2.0107$ & $2.3654$ & $-0.86909$ \\  
$-0.252$ & $-0.17$  & $10.020$  & $2.2432$ & $0.61679$   
                    & $-1.8783$ & $2.2432$ & $-0.83258$ \\  
\end{tabular}
\end{center}
\end{table}

\newpage

\begin{table}[h]
\begin{center}
\caption{\label{tab:vp} Finite parts $v_\diag^{\prime}$ 
and $v_\mix^{\prime}$ for
vertex correction of the baryon number violating operators.
Coefficients of the term $\csw^n$ $(n=0,1,2)$ are given in the 
column marked as $(n)$.}
\begin{tabular}{ll|lll|lll}
\multicolumn{2}{c}{gauge action} & 
\multicolumn{3}{c}{$v_\diag^{\prime}$} & 
\multicolumn{3}{c}{$v_\mix^{\prime}$} \\
$c_1$ & $c_{2}+c_{3}$ & $(0)$ & $(1)$ & $(2)$ & $(0)$ & $(1)$ & $(2)$ \\
\hline
$0$      & $0$      & $-25.821$ & $13.753$ & $-1.1792$   
                    & $-6.3894$ & $4.5991$ & $-1.1792$ \\  
$-1/12$  & $0$      & $-21.129$ & $11.779$ & $-1.0080$   
                    & $-5.0767$ & $3.8148$ & $-1.0080$ \\  
$-0.331$ & $0$      & $-14.319$ & $8.6222$ & $-0.72746$   
                    & $-3.2392$ & $2.6134$ & $-0.72746$ \\  
$-0.27$  & $-0.04$  & $-14.873$ & $8.8984$ & $-0.75215$   
                    & $-3.3803$ & $2.7111$ & $-0.75215$ \\  
$-0.252$ & $-0.17$  & $-13.583$ & $8.2648$ & $-0.69439$   
                    & $-3.0321$ & $2.4672$ & $-0.69439$ \\  
\end{tabular}
\end{center}
\end{table}

\newpage

\begin{table}[h]
\begin{center}
\caption{\label{tab:vpp} Finite parts $v_\diag^{\prime\prime}$ 
and $v_\mix^{\prime\prime}$ for
vertex correction of the baryon number violating operators.
Coefficients of the term $\csw^n$ $(n=0,1,2)$ are given in the 
column marked as $(n)$.}
\begin{tabular}{ll|lll|lll}
\multicolumn{2}{c}{gauge action} & 
\multicolumn{3}{c}{$v_\diag^{\prime\prime}$} & 
\multicolumn{3}{c}{$v_\mix^{\prime\prime}$} \\
$c_1$ & $c_{2}+c_{3}$ & $(0)$ & $(1)$ & $(2)$ & $(0)$ & $(1)$ & $(2)$ \\
\hline
$0$      & $0$      & $-1.3783$  & $1.1792$  & $-0.35763$   
                    & $4.5459$   & $-2.6297$ & $0.54935$ \\  
$-1/12$  & $0$      & $-1.1210$  & $1.0080$  & $-0.31313$   
                    & $3.5094$   & $-2.0995$ & $0.44985$ \\  
$-0.331$ & $0$      & $-0.74755$ & $0.72746$ & $-0.23554$   
                    & $2.1314$   & $-1.3427$ & $0.30008$ \\  
$-0.27$  & $-0.04$  & $-0.77933$ & $0.75215$ & $-0.24280$   
                    & $2.2182$   & $-1.3932$ & $0.31101$ \\  
$-0.252$ & $-0.17$  & $-0.71136$ & $0.69439$ & $-0.22656$   
                    & $1.9313$   & $-1.2279$ & $0.27829$ \\  
\end{tabular}
\end{center}
\end{table}

\end{document}